# Coupling MOS Quantum Dot and Phosphorus Donor Qubit Systems

M. Rudolph[1], P. Harvey-Collard[1,2], R. Jock[1], N.T. Jacobson[1], J. Wendt[1], T. Pluym[1], J. Dominguez[1], G. Ten-Eyck[1], R. Manginell[1], M.P. Lilly[1,3], M.S. Carroll[1]

[1]Sandia National Laboratories, Albuquerque, NM, USA, email: mrudolp@sandia.gov
[2]Université de Sherbrooke, Sherbrook, Québec, Canada
[3]Center for Integrated Nanotechnologies, Sandia National Labs, Albuquerque, NM, USA

*Abstract*—Si-MOS based QD qubits are attractive due to their similarity to the current semiconductor industry. We introduce a highly tunable MOS foundry compatible qubit design that couples an electrostatic quantum dot (QD) with an implanted donor. We show for the first time coherent two-axis control of a two-electron spin qubit that evolves under the QD-donor exchange interaction and the hyperfine interaction with the donor nucleus. The two interactions are tuned electrically with surface gate voltages to provide control of both qubit axes. Qubit decoherence is influenced by charge noise, which is of similar strength as epitaxial systems like GaAs and Si/SiGe.

## I. Introduction

Quantum computing has garnered significant attention due to the potential of significantly increasing computing efficiency. Si-MOS based quantum dot (QD) schemes are of particular interest due to their similarities to the mature technologies of the current semiconductor computing industry. Silicon also produces an effective "magnetic vacuum" through $^{28}$Si enrichment, which leads to high fidelity qubits. The longest coherence times in the solid-state have been demonstrated in electron and nuclear spins of donors in $^{28}$Si. However, direct coupling of two or more donors in Si has proven difficult to achieve. Coupling of quantum dots to one another, on the other hand, has been demonstrated, but their qubits do not achieve the same fidelities. We show for the first time two-axis control of a hybrid qubit coupling QD and donor systems. This produces a new, compact, all-electrical, potentially high speed, singlet-triplet two-electron qubit that is directly coupled to a second, nuclear spin qubit.

## II. Device Fabrication

The device active regions starts as a $^{28}$Si/SiO$_2$/poly-Si/SiN$_x$ stack grown on a Si handle. The layer thicknesses are 2.5 µm/35 nm/200 nm/35 nm (Figure 1(b)). The quantum dot structure is defined by electron beam lithography and selective dry etching of the poly-Si, which produces the gate pattern shown in Figure 1(a). Appropriate gate biases confine two quantum dots under gates (L)UCP, which are coupled to reservoirs under gates (L)URG and (L)ULG. An exploded schematic of the device highlights the Si/SiO$_2$ interface as presented in Figure 1(c). Phosphorus donors are implanted through a mask formed by a second EBL step. The poly-Si gate self-aligns the implant to the QD location, resulting in donors reaching two regions on the edge of the LCP gate (Figure 1(c)).

About 20 donors with energy 45 keV make it to each side of the gate, with a nominal depth of 25 nm and 15 nm of straggle. The straggle provides a finite probability that a donor will land underneath the LCP gate and also near the interface, which is important for coupling the QD and donor. For device operation, the QD under LCP is studied, while the QD under UCP is used to measure the charge state of the QD under LCP and donors. The device is probed by standard low-noise electrical methods in a dilution refrigerator at 10 mK.

## III. QD Properties

First, we study the electron transport through the QD under LCP, where source and drain are the reservoirs under LLG and LRG. For a well isolated QD, the Coulomb repulsion impedes the addition of an electron and separates the QD energy levels by a $E_C=e^2/C_{QD}$, with $C_{QD}$ the QD capacitance. Conduction is only possible when the QD chemical potential is resonant with the Fermi level, called Coulomb blockade. Figure 2(a) shows Coulomb blockade while scanning two gates, which maps out the charge stability of the QD. The stable periodicity of the Coulomb blockade peaks indicates the many electron regime. The diagonal break cutting through the resonances is the ionization event of a nearby donor electron.

The QD is emptied by making gates more negative to deplete the electrons. This action must be balanced with keeping the source/drain barriers open to still allow for conduction through the QD. Figure 2(b) shows a stability diagram when the QD is emptied completely, which is only possible with good individual control of the source and drain tunnel barriers. For non-optimized barriers, either conduction will be cut-off or energy levels will be broadened, both observed in Figure 2(b). The spacing between Coulomb blockade resonances becomes larger for fewer electrons as $C_{QD}$ decreases. To confirm the electron occupations annotated onto Figure 2(b), we measure Coulomb diamonds, where the gate voltage is scanned against the source/drain bias $V_{SD}$ (Figure 2(c)). As $V_{SD}$ increases, the Coulomb blockade region is decreased, disappearing for $V_{SD}> E_C$. For the empty QD, the diamond continues to open as there is no energy level left to tunnel through. The evolution of the chemical potential with an applied magnetic field elucidates the electron spin state (Figure 2(d,f)). N=0→1 only contains a single electron, so the chemical potential shifts with the Zeeman energy, which is monotonic in $B$. However, N=1→2 contains two electrons, and the competition between the Zeeman energy and the exchange energy produce a kink in the magnetospectroscopy when the spin state changes from singlet to polarized triplet.

While the conductance measurements can infer the charge occupation of the QD by counting Coulomb blockade resonances, it is not a direct measure of the charge state in the system. For this, a QD is formed under UCP that capacitively senses the charge occupation of nearby isolated objects. Figure 3(a) shows the response of the charge sensor in the few-electron regime of the QD under LCP. The broad features are the background signal of the charge sensor itself, while the charge transitions are the sharp peaks that have been labeled. Isolated donor electrons are visible with this technique, which does not require direct transport. The QD transitions are observed as parallel lines, while at least three donor transitions are also present. Figure 3(b) shows the magnetospectroscopy in charge sensing and supports the electron occupations in Figure 3(a). The N=1→2 kink in magnetospectroscopy is a measure of the Si valley splitting $E_{VS}$ in the QD. In Figure 3(c), we show that we can tune $E_{VS}$ by increasing the vertical electric field with gate LCP. Valley splittings large enough to provide well separated energy levels are attainable (electron temperature is 170 mK). This is an important demonstration, as small valley splitting is a detriment to electron coherence. Note that the N=3→4 transition has an even larger exchange splitting arising from the orbital energy levels, which is leveraged for the qubit measurements. Figure 3(d) displays the tunability of the QD-reservoir tunnel rate $\Gamma$ with gate LLG, showing about 1 decade per 100 mV. The QD energy shift for 100 mV on LLG is only 1.9 meV (~$0.2E_C$), indicating good orthogonal control between the tunnel rate and the charge occupation.

## IV. QD-DONOR QUBIT

To perform coherent electron spin rotations in a QD-donor system, we operate in an effective 2-electron singlet/triplet (S/T) space. Figure 4(a) shows the (3,1)-(4,0) (QD,donor) charge transition, where the first two QD electrons are in a closed shell in the QD. We pulse from (3,0) to (4,0) to initialize a mixture of S and T in the QD. Next, we pulse point M to every pixel on the figure to measure which state was loaded. The (4,0)-(3,1) resonance is shifted between S and T by the exchange energy. Both states are observed since a mixture is loaded. The region between the two resonances is a Pauli spin blockade region, where the excited state (T) lifetime can be 100s of microseconds, and can be used to measure the spin state of the electron.

Coherent rotations of the electron spin are achieved by deterministically loading a (4,0)S and then moving one electron onto the donor. The electron interacts with the donor nucleus via the hyperfine interaction while the electron that remains on the QD does not. This effective magnetic field gradient induces a relative spin rotation between the two electrons. A relative phase accumulates for the amount of time that the electrons are separated. Figure 4(b) shows the probability that a T is measured after separating the electrons for a manipulation time. The x-axis changes how far the QD and donor chemical potentials are detuned, with ε=0 being resonance. For ε<0, the electrons are never separated and S is always measured. For ε>0, coherent S/T rotations are observed. Figure 4(c) extracts the rotation frequency at different detuning points, ε. For large ε, the qubit rotates by the hyperfine interaction, which is relatively constant in ε. For small ε, the frequency increases significantly, as the exchange interaction between the electrons dominates. This demonstrates control of two orthogonal qubit axes, important for universal gate control.

Figure 4(d) displays the rotations for select detuning points, indicating that the increase in exchange correlates with the decrease in coherence, indicating the influence of charge noise in the qubit. The decay envelope fits well to a Gaussian, supporting a quasi-static charge noise spectrum. Plotting the characteristic decay time $T_2^*$ against ε shows a $\varepsilon^2$ dependence for small ε, consistent with a quasi-static spectrum with a magnitude similar to GaAs and Si/SiGe systems. The saturation of $T_2^*$ for large ε is likely due to magnetic noise from background $^{29}$Si nuclei.

## V. CONCLUSIONS

We have developed a highly tunable MOS foundry compatible QD device that is capable of achieving a one-electron QD that is coupled to an intentionally implanted donor. We show coherent spin rotations about two qubit axes defined by this new hybrid QD-donor qubit system using exchange coupling and the donor contact hyperfine coupling to the nuclear spin qubit's magnetic polarization. Charge noise of the MOS system is characterized and is comparable to other semiconductor systems indicating that MOS is a viable interface for ST qubits. The QD-donor qubit can be extended by using the donor nuclear spin information and coupling multiple QD-donor cells to scale out to multi-qubits.

ACKNOWLEDGMENT

This work was performed, in part, at the Center for Integrated Nanotechnologies, a U.S. DOE, Office of Basic Energy Sciences user facility. This work was supported by the Laboratory Directed Research and Development program at Sandia National Laboratories. Sandia National Laboratories is a multi-program laboratory managed and operated by Sandia Corporation, a wholly owned subsidiary of Lockheed Martin Corporation, for the U.S. Department of Energy's National Nuclear Security Administration under contract DE-AC04-94AL85000.

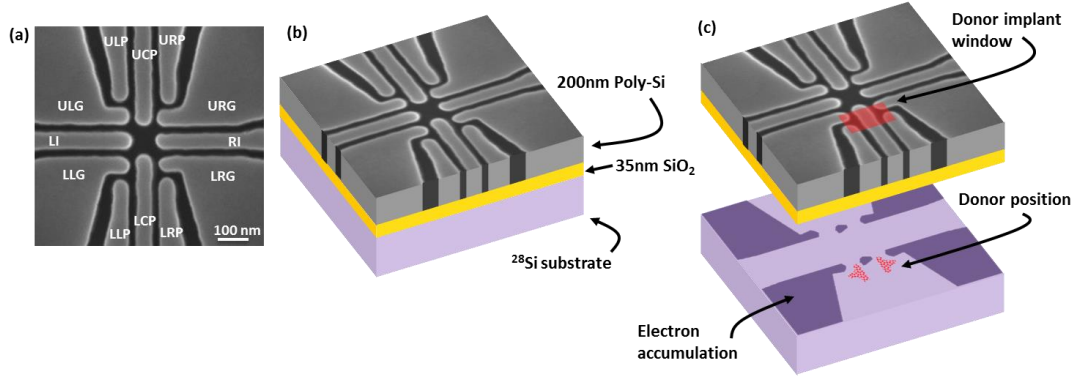

Fig. 1 (a) Top-down device scanning electron micrograph of the poly-Si gates (light gray). (b) Schematic of the $^{28}$Si MOS stack with 35 nm gate oxide. (c) Exploded schematic indicated the regions of electron accumulation at the interface during operation, during which two quantum dots are formed. Phosphorus donors are implanted through the mask window indicated by the red rectangle. The poly-Si further masks the incident donors, and the approximate locations of the donors that stop in the substrate are indicated by the red dots.

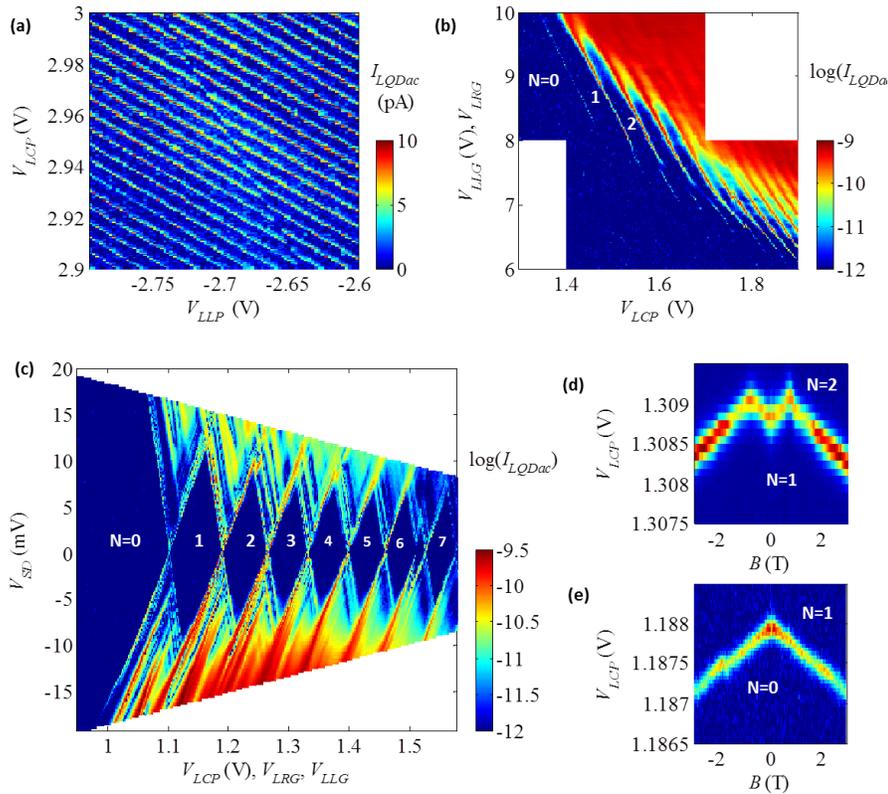

Fig. 2 (a) Charge stability map showing Coulomb blockade resonances of a many electron quantum dot under LCP. (b) Charge stability map of the few electron regime, with the empty dot state (N=0) regime indicated. (c) Coulomb diamond map of the few electron regime with indicated charge occupations. (d,e) Magnetospectroscopy of the N=1→2 and N=0→1 transitions, indicating a single spin state for the one electron dot and a singlet/triplet splitting of 1T for the two electron dot.

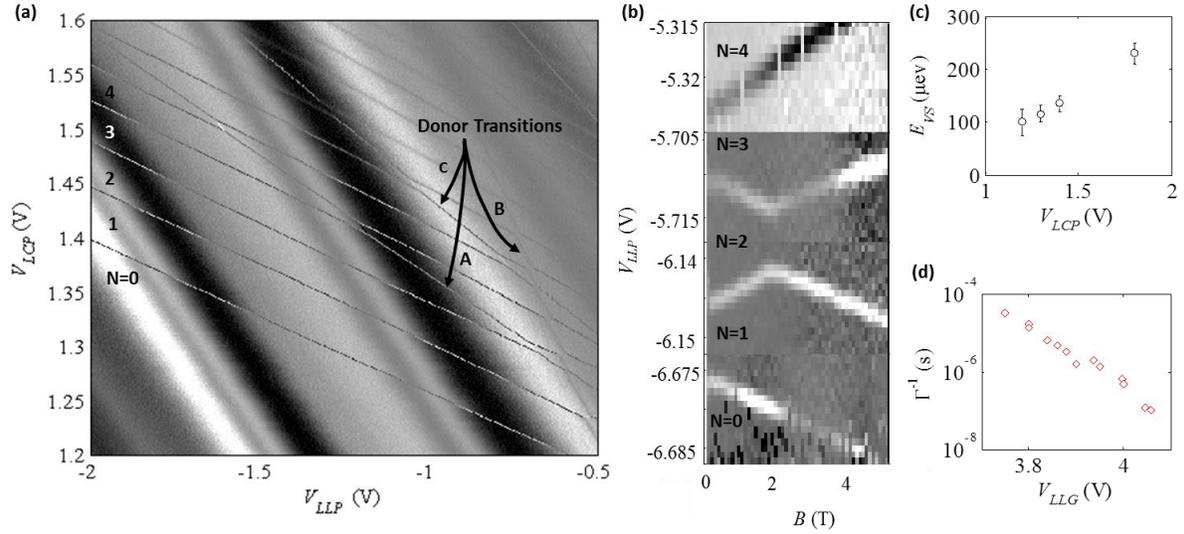

Fig. 3 (a) Conductance response of the quantum dot under UCP sensing charge transitions for the few electron dot under LCP. Three charge transitions for donors are indicated, which have a different slope than the few electron dot. (b) Magnetospectroscopy of the first four transitions in the quantum dot. (c) The valley splitting as extracted from the magnetospectroscopy for different values of LCP. LCP locally controls the vertical electric field at the quantum dot, and thus has a significant effect on the valley splitting. (d) Tunnel time of an electron from the reservoir onto the quantum dot for values of LLG.

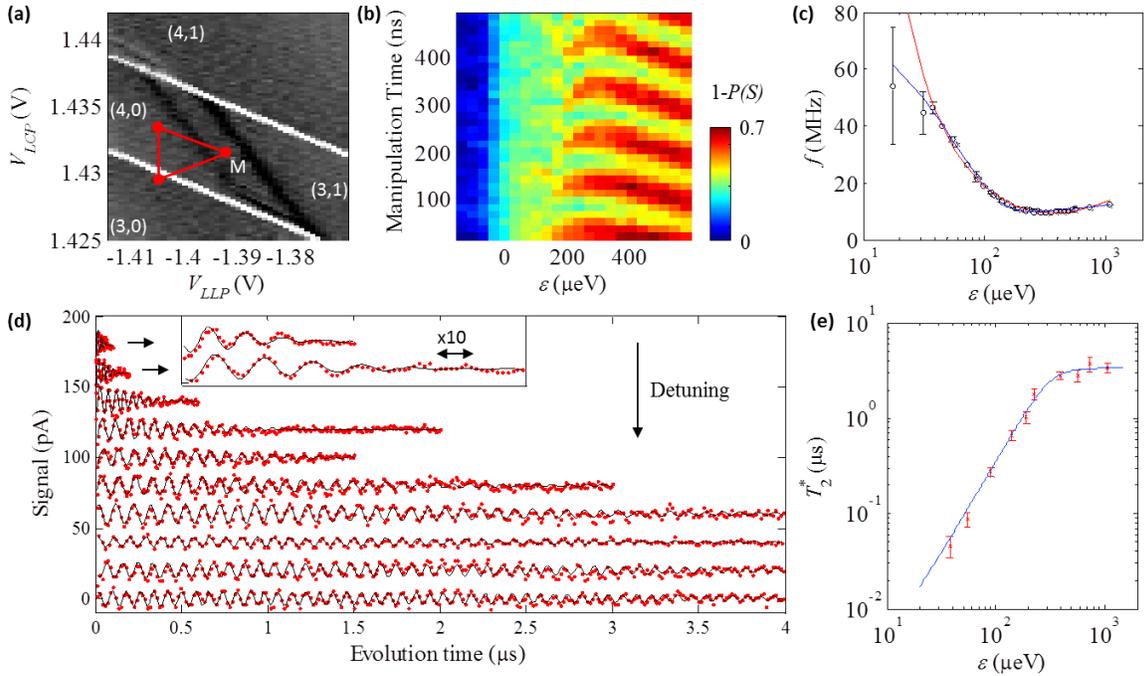

Fig. 4 (a) Stability diagram at the (4,0)-(3,1) charge transition after loading a mixture of the (4,0)S and (4,0)T states (red pulsing trajectory the measurement level M scanned across the voltage range). Both the S and T anticrossings are visible due to the Pauli blockade mechanism. (b) Rabi oscillations for different manipulation detunings ε. (c) Oscillation frequency as a function of ε showing a sharp increase in the exchange gap for small ε, with fits to a power law (red) and exponential (blue) dependence. (d) Coherence of the rotations for different ε (inset expands the time-axis for the two smallest ε for clarity). The fits are to a Gaussian decay envelope. (e) Extracted $T_2^*$, showing a $ε^2$ dependence for small ε and a saturation for large ε.